\documentclass[12pt]{article}

\usepackage{amssymb}
\usepackage{amsopn}
\usepackage{graphicx}

\def\hybrid{\topmargin 0pt	\oddsidemargin 0pt 
	\headheight 0pt	\headsep 0pt
	\textheight 9in		
	\textwidth 6.1in	
	\marginparwidth .875in
	\parskip 5pt plus 1pt	\jot = 1.5ex}

\hybrid

\catcode`\@=11

\def\numberbysection{\@addtoreset{equation}{section}
	\def\theequation{\thesection.\arabic{equation}}}

\def\titlepage{\@restonecolfalse\if@twocolumn\@restonecoltrue\onecolumn
     \else \newpage \fi \thispagestyle{empty}\c@page\z@	
	\def\thefootnote{\fnsymbol{footnote}} }

\def\endtitlepage{\if@restonecol\twocolumn \else \newpage \fi
	\def\thefootnote{\arabic{footnote}} 
	\setcounter{footnote}{0}}  

\catcode`@=12
\relax

\numberbysection

\def\mysavedown#1{\edef\mysubs{\mysubs#1}}
\def\mysaveup#1{\edef\mysups{\mysups#1}}
\def\mydown#1{{\mytensor}_{\vphantom{\mysubs}#1}}
\def\myup#1{{\mytensor}^{\vphantom{\mysups}#1}}
\def\tensor#1#2{
  #1
  \def\mytensor{\vphantom{#1}}
  \def\mysubs{\relax}
  \def\mysups{\relax}
  \let\down=\mysavedown
  \let\up=\mysaveup
  #2
  \let\down=\mydown
  \let\up=\myup
  #2
  }

\DeclareMathOperator{\GL}{GL}

\DeclareMathOperator{\OrthQ}{O_{Q}}
\DeclareMathOperator{\Orth}{O}
\DeclareMathOperator{\SOrth}{SO}
\DeclareMathOperator{\Sp}{Sp}
\DeclareMathOperator{\Tr}{Tr}

\DeclareMathOperator{\soQ}{\mathfrak{s}\mathfrak{o}_{Q}}
\DeclareMathOperator{\so}{\mathfrak{s}\mathfrak{o}}
\newcommand{\Btil}{\widetilde{B}}

\newcommand{\Gtil}{\widetilde{G}}
\newcommand{\Htil}{\widetilde{H}}
\newcommand{\Mtil}{\widetilde{M}}

\newcommand{\Pitil}{\widetilde{\Pi}}
\newcommand{\Ptil}{\widetilde{P}}
\newcommand{\Rtil}{\widetilde{R}}
\newcommand{\Ttil}{\widetilde{T}}
\newcommand{\Vtil}{\widetilde{V}}
\newcommand{\Wtil}{\widetilde{W}}

\newcommand{\antisym}{\mathfrak{A}}

\newcommand{\bbR}{\mathbb{R}}

\newcommand{\etil}{\tilde{e}}

\newcommand{\ftil}{\tilde{f}}
\newcommand{\graph}{\Gamma_{\! f}}
\newcommand{\gtil}{\tilde{g}}
\newcommand{\half}{\frac{1}{2}}

\newcommand{\liegtil}{\tilde{\mathfrak{g}}}
\newcommand{\lieg}{\mathfrak{g}}

\newcommand{\ltil}{\tilde{l}}

\newcommand{\mixed}{\mathfrak{M}}

\newcommand{\mtinv}{(m^{t})^{-1}}
\newcommand{\myLie}[1]{\mathcal{L}_{#1}}
\newcommand{\ntil}{\tilde{n}}

\newcommand{\omegatil}{\tilde{\omega}}
\newcommand{\pitil}{\tilde{\pi}}

\newcommand{\ptil}{\tilde{p}}
\newcommand{\qtil}{\tilde{q}}

\newcommand{\thetatil}{\tilde{\theta}}

\newcommand{\xstil}[1]{\xtil^{#1}{}_{\sigma}}
\newcommand{\xs}[1]{x^{#1}{}_{\sigma}}
\newcommand{\xtil}{\tilde{x}}

\newcommand{\oaref}[2]{II-#2}

\begin{document}
\bibliographystyle{utphys}
\begin{titlepage}
\noindent
\strut\mbox{March 2000}\hfill{UMTG--221}\newline
\strut\hfill{\tt hep-th/0003177}
\par\vskip 2cm
\begin{center}
    {\Large \bf Target Space Duality I: General Theory\footnote{This work
    was supported in part by National Science Foundation grant
    PHY--9870101.}}\\[0.5in]
    {\bf Orlando Alvarez}\footnote{email: \tt oalvarez@miami.edu}\\[0.1in]
    Department of Physics\\
    University of Miami\\
    P.O. Box 248046\\
    Coral Gables, FL 33124
\end{center}
\par\strut\vspace{.5in}
\noindent

\begin{abstract}
    We develop a systematic framework for studying target space duality at
    the classical level.  We show that target space duality between
    manifolds $M$ and $\Mtil$ arises because of the existence of a very
    special symplectic manifold.  This manifold locally looks like
    $M\times\Mtil$ and admits a double fibration.  We analyze the local
    geometric requirements necessary for target space duality and prove
    that both manifolds must admit flat orthogonal connections. We 
    show how abelian duality, nonabelian duality and Poisson-Lie 
    duality are all special cases of a more general framework. As an 
    example we exhibit new (nonlinear) dualities in the case $M=\Mtil=\bbR^{n}$.
\end{abstract}

\vspace{.5in}

\end{titlepage}

\section{Introduction}
\label{sec:introduction}

The $(1+1)$ dimensional sigma model describes the motion of a string
on a manifold.  The sigma model is specified by giving a triplet of data
$(M,g,B)$ where $M$ is the target $n$-dimensional manifold, $g$ is a
metric on $M$, and $B$ is a $2$-form on $M$.  The lagrangian for this
model is
\begin{equation}
    \mathcal{L} = \half g_{ij}(x)
    \left(
    \frac{\partial x^{i}}{\partial \tau}
    \frac{\partial x^{j}}{\partial \tau}
    -\frac{\partial x^{i}}{\partial \sigma}
    \frac{\partial x^{j}}{\partial \sigma}
    \right) +
    B_{ij}(x)
    \frac{\partial x^{i}}{\partial \tau}
    \frac{\partial x^{j}}{\partial \sigma}
    \label{eq:lag}
\end{equation}
with canonical momentum density
\begin{equation}
    \pi_{i}= \frac{\partial\mathcal{L}}{\partial \dot{x}^{i}} =
    g_{ij}\dot{x}^{j} + B_{ij}x'^{j}\;,
    \label{eq:canmomentum}
\end{equation}
where an overdot denotes the time derivative ($\partial/\partial\tau$)
and a prime denotes the space derivative ($\partial/\partial\sigma$)
on the worldsheet.  What is remarkable is that it possible for two
completely different sigma models, $(M,g,B)$ and
$(\Mtil,\gtil,\Btil)$, to describe the same physics.  By this we mean
that there is a canonical transformation between the space of paths on
$M$ and the corresponding one on $\Mtil$ that preserves the respective
hamiltonians.  This phenomenon is known as \emph{target space
duality}.

This is the first of two articles where we develop a systematic
framework for studying target space duality at the classical level. 
We do not consider quantum aspects of target space duality nor do we
consider examples involving mirror symmetry.  Most of our
considerations are local but phrased in a manner that is amenable to
globalization.  We analyze the local geometric requirements necessary
for target space duality.  The study of target space duality has
developed by discovering a succession of more and more complicated
examples (see below).  We show that the known examples of abelian
duality, nonabelian duality and Poisson-Lie duality are all derivable
as special cases of the framework.  We show that target space duality
boils down to the study of some very special symplectic manifolds that
allow the reduction of the structure group of the frame bundle 
to $\SOrth(n)$.  In article~I we develop the general theory
and apply it so some very simple examples.  In 
article~II~\cite{Alvarez:2000bi} we
systematically apply the theory to a variety of scenarios and we
reproduce nonabelian duality and Poisson-Lie duality.  
The theory is applied to other geometric situations that lead us deep into
unknown questions in Lie algebra theory. We try to make article~I self 
contained. References to equations and sections in article~II are 
preceded by~II, \emph{e.g.}, (II-8.3).

What is the value in developing a general framework for studying
classical target space duality?  The framework may say something about
the what is string theory.  We believe that there is some parameter
space that describes string theory.  For special values of the
parameters we get the familiar Type I, Type II-A, Type II-B,
\emph{etc}.  theories and that these are related by various dualities. 
If we can get a handle on the class of symplectic manifolds that lead
to target space duality we may be able to get a better idea about the
parameter space of string theory.

The simplest target space duality is abelian duality.  Here a theory
with target space $S^{1}$ or $\bbR$ is dual to a theory with target
space $S^{1}$ or $\bbR$.  For a comprehensive review and history of
abelian duality look in \cite{Giveon:1994fu}.  It should also be
mentioned that it has been known for a long time, see \emph{e.g.}
\cite{Giveon:1989tt}, that the abelian duality transformation is a
canonical transformation.  A first attempt to generalize abelian
duality to groups led to the pseudochiral model of Zakharov and
Mikhailov \cite{Zakharov:1978pp} as a dual to the nonlinear sigma
model.  Nappi \cite{Nappi:1980ig} showed that these models were not
equivalent at the quantum level.  The correct dual model was first
found by Fridling and Jevicki~\cite{Fridling:1984ha} and Fradkin and
Tseytlin~\cite{Fradkin:1985ai} using path integral methods.  String
theory motivated a renewed interest in abelian and nonabelian duality
\cite{Kiritsis:1991zt,Rocek:1992ps,Giveon:1992jj,delaOssa:1993vc,%
Gasperini:1993nz,Giveon:1994mw,Giveon:1994ai,Giveon:1994ph}.  
It was shown that the
duality transformation was canonical
\cite{Curtright:1994be,Alvarez:1994wj} and these ideas were
generalized in a variety of ways 
\cite{Alvarez:1994zr,Alvarez:1994qi,Lozano:1995jx,Lozano:1996sc,%
Alvarez:1995uc}. The form of the generating functions for duality 
transformation gave hints that nonabelian duality was
associated with the geometry of the cotangent bundle of the group.

The most intricate target space duality discovered thus far is the
Poisson-Lie duality of Klimcik and Severa
\cite{Klimcik:1995ux,Klimcik:1996dy,Klimcik:1996kw}.  In this example
we see a very nontrivial geometrical structure playing a central role. 
A Poisson-Lie group $G$ is a Lie group with a Poisson bracket that is
compatible with the group multiplication law.  Drinfeld
\cite{Drinfeld:83aa} showed that Poisson-Lie groups are determined 
by a Lie bialgebra $\lieg_{D}= \lieg\oplus\liegtil$ where $\lieg$ is
the Lie algebra of $G$ and $\liegtil$ is the Lie algebra of a Lie
group $\Gtil$, See Appendix~\oaref{sec:bialgebra}{B.1}.  The two Lie algebras
are coupled together in a very symmetric way.  A Lie group $G_{D}$
with Lie algebra $\lieg_{D}$ is called a Drinfeld double.  It should
be pointed out that $\Gtil$ is also a Poisson-Lie group.  By using a
clever argument, Klimcik and Several discovered that if the metric $g$
and $B$ field on a Poisson-Lie group $G$ was of a special form then
there would be a corresponding metric $\gtil$ and $\Btil$-field on the
group $\Gtil$.  Their observations follow from the symmetric way that
$G$ and $\Gtil$ enter into the Drinfeld double $G_{D}$.  They showed
that that by writing down a ``first order'' sigma model on $G_{D}$
they could derive either the model on $G$ or the model on $\Gtil$ by
taking an appropriate slice.  Here one explicitly sees that the the
target manifold and the target dual manifold are carefully glued
together into a larger space.  Klimcik and Severa do not explicitly
write down the duality transformation but they are totally explicit
about the metric and $B$ field.  It was Sfetsos
\cite{Sfetsos:1998pi,Sfetsos:1996xj} who wrote down the duality
transformation, verified that it was a canonical transformation, and 
constructed the generating function for the canonical transformation,
see also \cite{Stern:1998my}.

At the time of the work by Klimcik and Severa, the author had been
working on a program to develop a general theory of target space
duality, see \cite{Alvarez:1995hd}.  In that article I advocated the use
of generating functions of the type (\ref{eq:defF}) because they would
lead to a linear relationship\footnote{For nonpolynomial generating 
functions look at \cite{Sfetsos:1999zm}.}
between $(dx/d\sigma,\pi)$ and
$(d\xtil/d\sigma,\pitil)$ that preserved the quadratic nature of the 
sigma model hamiltonians.  I discussed the geometry which was
involved and explained the role played in this geometry by the
hamiltonian density $\mathcal{H}$ and the momentum density
$\mathcal{P}$.  Explicit formulas relating the geometries of the two
manifolds were not given in that article for the following reason. 
The formulation I had at the time involved variables $(x,p)$ where
essentially $\pi=dp/d\sigma$.  This gave a certain symmetry to some of
the equations but at a major price.  The $B$ field gauge symmetry
$B\to B+dA$ became a nonlocal symmetry in $(x,p)$ space and the gauge
symmetry was no longer manifest.  Only for special choices of $A$ was
the gauge transformation local.  The formulas I had derived respected
the special gauge transformations but I could not verify general gauge
invariance.  Sfetsos~\cite{Sfetsos:1998pi} exploited some of the
geometric constraints I had proposed and he was able to explicitly
construct the duality transformation for Poisson-Lie duality. 
Sfetsos' work is very interesting.  He conjectures the form of the
duality transformation and he knows the geometric data $(M,g,B)$ and
$(\Mtil,\gtil,\Btil)$ from the work of Klimcik and Severa.  He now
uses this information and certain integrability constraints to
explicitly work out the generating function for the canonical
transformation.  Sfetsos' computation may be reinterpreted
as the construction of a known symplectic structure
\cite{Alekseev:1994qs,Alekseevsky:98aa} on the Drinfeld double, see
Section~\oaref{sec:Poisson-Lie}{3}.

In this article I present a general theory for target space duality
that is manifestly gauge invariant with respect to $B$ field gauge
transformations.  I consider what could be called \emph{irreducible
duality} where there are no spectator fields.  All the fields
participate actively in the duality transformation.  I show that the
duality transformation arises because of the existence of a special
symplectic manifold $P$ that locally looks like $M\times\Mtil$ and
admits a double fibration.  The duality transformation exists only
when there exists a compatible confluence of several distinct
geometric structures associated to the manifold $P$: an $\Orth(2n)$
structure related to the hamiltonian density (\ref{eq:hamiltonian}),
an $\Orth(n,n)$ structure related to the momentum density
(\ref{eq:momentum}), an $\Orth(n)\times\Orth(n)$ structure associated
with the sigma model metrics, and a $\Sp(2n)$ structure related to the
symplectic form.  This is why these symplectic manifolds are very
special and rare.  I develop the general theory and then show how the
known examples of abelian duality, nonabelian duality and Poisson-Lie
duality follow.  The general theory indicates that there are probably
many more examples.  For example, in Section~\ref{sec:ftilzero} I
write down families of nonlinear duality transformations that map a
theory with target space $\bbR^{n}$ into one with target space
$\bbR^{n}$.  I also investigate a variety of scenarios and pose open
mathematical questions deeply related to the theory of Lie
algebras.

This work differs from the work of Sfetsos~\cite{Sfetsos:1998pi} in a
variety of ways.  There are two types of constraints on the canonical
transformation: algebraic constraints having to do with quadratic form
of the hamiltonian density and differential constraints having to do
integrability conditions.  Sfetsos writes these down but in a way that
is neither geometric nor gauge invariant.  He applies them to
Poisson-Lie duality and derives the generating function.  Sfetsos'
formulation does not exploit the fact that there are natural geometric
structures associated to these equations.  This is what I was trying
to do in \cite{Alvarez:1995hd} but failed due to a bad choice of
variables $(x,p)$ leading to an absence of manifest $B$ field gauge
invariance.  The formulation presented here uses the variables
$(x,\pi)$ and is manifestly gauge invariant.  In
Section~\oaref{sec:noncartesian}{2.2.2} I give a geometric interpretation of
$B$ field gauge invariance.  In this article I work in terms of
adapted frame fields.  In this way, the formalism has an immediate
interpretation in terms of $H$-structures on the bundle of frames.  In
fact the discussion presented in Section~\oaref{sec:inf-hom-n}{4.1} is done
in a sub-bundle of the bundle of frames.

The framework developed in this work allows one to attack a variety of
interesting questions.  Are there any interesting restrictions on the
manifolds $M$ and $\Mtil$?  We show in
Section~\ref{sec:algebraicconstraints} that the manifolds $M$ and
$\Mtil$ have to admit flat orthogonal connections.  We know for any
manifold $M$ there always exists a natural symplectic manifold
$P=T^{*}M$, the cotangent bundle.  We can ask what type of dualities
arises from the standard symplectic structure on the cotangent bundle? 
We show that this can only happen if $M$ is a Lie group, see
Section~\oaref{sec:cotangent}{2.2.1}.  This formalism allows general question
to be asked.  For example there are a series of PDEs that have to be
solved to determine the duality transformations.  These PDEs depend on
some functions.  If these functions are zero then one gets abelian
duality, if some are made nonzero then you get nonabelian duality,
\emph{etc}.  This is a framework that can be used for a systematic study of
duality. It opens up the possibility to study dualities involving 
parallelizable manifolds that are not Lie groups such as $S^{7}$ or 
sub-bundles of the frame bundle. This work indicates that duality is 
a very rich geometrical framework ripe for study and we have only 
scratched the surface.

\section{The symplectic structure}
\label{sec:symplectic}

We review briefly the notion of a ``generating function'' in canonical 
transformations because our methods introduce a secondary symplectic 
structure into the formulation of target space duality and it is 
important to understand the difference between the two.

Assume you have symplectic manifolds, $P$ and $\Ptil$, with respective
symplectic forms $\omega$ and $\omegatil$.  Consider $P\times\Ptil$
with standard projections $\Pi:P\times\Ptil\to P$ and
$\Pitil:P\times\Ptil\to\Ptil$.  You can make $P\times\Ptil$ into a
symplectic manifold by choosing as symplectic form
$\Omega=\Pitil^{*}\omegatil-\Pi^{*}\omega$.  By definition, a
canonical or symplectic transformation $f:P\to\Ptil$ satisfies
$f^{*}\omegatil=\omega$.  We describe $f$ by its graph $\graph\subset
P\times\Ptil$.  It is clear $f:P\to\Ptil$ will be symplectic if and
only if $\Omega|_{\graph}=0$.  Locally we have $\omega=d\theta$ and
$\omegatil=d\thetatil$.  Thus we see that $\thetatil-\theta$ is a
closed $1$-form on $\graph$.  Consequently there exists locally a
function $F:\graph\to\bbR$ such that $\thetatil-\theta=dF$.  This
function $F$ is called the ``generating function'' for the symplectic
transformation.  The reason is that if in local Darboux coordinates we
have that $\theta = pdq$ and $\thetatil=\ptil d\qtil$ then we have
that $F$ is locally a function of only $q$ and $\qtil$, $\ptil =
\partial F/\partial\qtil$ and $p = -\partial F/\partial q$.  We can
now use the inverse function theorem to construct the map from $(q,p)$
to $(\qtil,\ptil)$.  Note that $\dim\graph=2n$ and therefore $F$ is a
function of $2n$ variables.  Had we chosen $\theta= -qdp$ then we would
have that $F$ is a function of $\qtil$ and $p$.  In this case it is
worthwhile to observe $F=\qtil p$ generates the identity
transformation.  We mention this because the identity transformation
is not in the class of transformations generated by functions of $q$
and $\qtil$.

All this generalizes to field theory.  We discuss only the case of 
$(1+1)$ dimensions.  Let $P(M)$ be the 
{\em path space} of M. By this we mean the set of maps $\{\gamma: N\to 
M\}$ where $N$ can be $\bbR$, $S^{1}$ or $[0,\pi]$ depending on 
whether we are discussing infinite strings, closed strings or open 
strings.  Most of the discussion in this article is local and so we do 
not specify $N$.  In the case of a sigma model with target space $M$, 
the basic configuration space is $P(M)$ with associated phase space 
$P(T^{*}M)$. If $(x,\pi)$ are coordinates on $T^{*}M$ then the 
symplectic structure on $P(T^{*}M)$ is given by
\begin{equation}
    \int \delta\pi(\sigma)\wedge\delta x(\sigma)\; d\sigma\;.
    \label{eq:symplectic}
\end{equation}
In what follows we are interested in looking for canonical 
transformations between a sigma model with target space $M$ and one 
with target space $\Mtil$ of the same dimensionality. We say that a 
sigma model with geometrical data $(M,g,B)$ is dual to a sigma 
model $(\Mtil,\gtil,\Btil)$ if there exists a canonical 
transformation $F: P(T^{*}M) \to P(T^{*}\Mtil)$ that preserves the 
hamiltonian densities, $F^{*}\widetilde{\mathcal{H}} =\mathcal{H}$, 
where the hamiltonian density is given by (\ref{eq:hamiltonian}).

In the case of ``abelian duality'' where the target space is a circle 
you can choose the generating function to be
$$
    F[x,\xtil] = \int \xtil \frac{dx}{d\sigma}\;d\sigma\;.
$$
This leads to the standard duality relations $\pi(\sigma) = 
d\xtil/d\sigma$ and $\pitil(\sigma)= dx/d\sigma$.

The nonabelian duality relations follow from the following natural
choice \cite{Lozano:1995jx,Alvarez:1995uc} for generating function. 
Assume the target space is a simple connected compact Lie group $G$
with Lie algebra $\mathfrak{g}$.  The dual manifold is the Lie algebra
with an unusual metric.  The generating function is very natural:
$$
    F[g,\widetilde{X}] = \int \Tr\left( \widetilde{X}\, 
    g^{-1}\frac{dg}{d\sigma}\right)d\sigma\;,
$$
where $\widetilde{X}$ is a Lie algebra valued field.

We now consider a class of generating functions for target space
duality that leads to a linear relationship \cite{Alvarez:1995hd}
between $(dx/d\sigma, \pi(\sigma))$ and the corresponding variables on
the dual space.  On $M\times\Mtil$ choose locally a $1$-form $\alpha=
\alpha_{i}(x,\xtil)dx^{i} + \tilde{\alpha}_{i}(x,\xtil)d\xtil^{i}$. 
We can define a natural ``generating function'' on $P(M\times\Mtil)$
by
\begin{equation}
    F[x(\sigma),\xtil(\sigma)] = \int\alpha = \int 
    \left(\alpha_{i}(x(\sigma),\xtil(\sigma))\frac{dx^{i}}{d\sigma} + 
    \tilde{\alpha}_{i}(x(\sigma),\xtil(\sigma)) 
    \frac{d\xtil^{i}}{d\sigma}\right)d\sigma\;.
    \label{eq:defF}
\end{equation}
We only consider target space duality that arises from this type of 
canonical transformation.

Let $v$ be a vector field along the path $(x(\sigma),\xtil(\sigma)) 
\in M\times\Mtil$ with compact support which represents a deformation 
of the path.  Note that $\delta_{v}F = \int \myLie{v}\alpha = \int 
\iota_{v}d\alpha$.  In the previous formula $\myLie{v}\alpha = 
\iota_{v}d\alpha + d\iota_{v}\alpha$ is the Lie derivative with 
respect to $v$.  Since $v$ has compact support, the exact term can be 
neglected.  Thus the variation of $F$ is determined by the exact 
$2$-form $\beta=d\alpha$:
\begin{equation}
    \delta_{v}F = \int \iota_{v}\beta\;.
    \label{eq:varF}
\end{equation}
We use $\beta$ to construct the duality transformation.  If $x$ and 
$\xtil$ are respectively local coordinates on $M$ and $\Mtil$ then
\begin{equation}
    \beta = -\half l_{ij}(x,\xtil) dx^{i}\wedge dx^{j} + 
    m_{ij}(x,\xtil) d\xtil^{i}\wedge dx^{j} + \half 
    \ltil_{ij}(x,\xtil) d\xtil^{i}\wedge d\xtil^{j}\;,
    \label{eq:beta}
\end{equation}
where $\ltil$: $l_{ij}=-l_{ji}$ and $\ltil_{ij}=-\ltil_{ji}$.  The
three $n\times n$ matrix functions $l,\ltil,m$ are used to construct
the canonical transformation on the infinite dimensional phase space. 
A brief calculation shows that the canonical transformations are
\begin{eqnarray}
    \pi_{i}(\sigma) & = & m_{ji}(x,\xtil)\frac{d\xtil^{j}}{d\sigma}
    	+l_{ij}(x,\xtil)\frac{dx^{j}}{d\sigma}\;,
    \label{eq:dualpi}  \\
    \pitil_{i}(\sigma) & = & m_{ij}(x,\xtil)\frac{dx^{j}}{d\sigma}
    	+ \ltil_{ij}(x,\xtil)\frac{d\xtil^{j}}{d\sigma}\;.
    \label{eq:dualpitil}
\end{eqnarray}
The invertibility of the canonical transformation between $P(T^{*}M)$ 
and $P(T^{*}\Mtil)$ requires $m$ to be an invertible matrix.  This 
implies that $\beta$ is of maximal rank, \emph{i.e.} a symplectic 
form\footnote{\label{foot:symplectic} It is possible for $\beta$ to be 
symplectic and have $m=0$ but this will not define an invertible 
canonical transformation between $P(T^{*}M)$ and $P(T^{*}\Mtil)$.  For 
example, if $M$ and $\Mtil$ are symplectic manifolds with respective 
symplectic forms $\omega$ and $\omegatil$ then choose 
$\beta=\omegatil-\omega$.} on $M\times\Mtil$.

It is important to recognize that there are two very different
symplectic structures in this problem.  The first one is the standard
symplectic structure on phase space $P(T^{*}M)$ given by
(\ref{eq:symplectic}).  The second one on $M\times\Mtil$ given by
$\beta$ arises from the class of generating functions (\ref{eq:defF})
we are considering.  The generating function arguments are local and
suggest that the symplectic structure on $M\times\Mtil$ may be
generalized to a symplectic manifold $P$ which ``contains''
$M\times\Mtil$.  In the cartesian product $M\times\Mtil$ you have
natural cartesian projections $\Pi_{c}:M\times\Mtil \to M$ and
$\Pitil_{c}:M\times\Mtil \to\Mtil$.  The product structure can be
generalized by the introduction of the concept of a bifibration.  A
$2n$ dimensional manifold $P$ is said to be a \emph{bifibration} if
there exists $n$ dimensional manifolds $M$ and $\Mtil$ and projections
$\Pi:P\to M$ and $\Pitil:P\to\Mtil$ such that the respective fibers
are diffeomorphic to coverings spaces of $\Mtil$ and $M$ and they are
also transverse.  This means that if $p\in P$ then $\ker \Pi_{*}|_{p}
\oplus \ker \Pitil_{*}|_{p} = T_{p}P$ where $\Pi_{*}$ and $\Pitil_{*}$
are the differential maps of the projections.  Note that the cartesian
product manifold $P = M\times\Mtil$ is an example of a bifibration. 
If the product projection $\Pi\times\Pitil:P\to M\times\Mtil$ is
injective\footnote{The definition of a fiber bundle implies that
$\Pi\times\Pitil$ is surjective.} then $P=M\times\Mtil$.  A covering
space example is given by $P=\bbR^{2}$ and $M=\Mtil=S^{1}$ with
$\Pi:(x,\xtil)\mapsto e^{ix}$ and $\Pitil:(x,\xtil)\mapsto
e^{i\xtil}$.

\begin{figure}[tbp]
    \centering
    \includegraphics[width=0.5\textwidth]{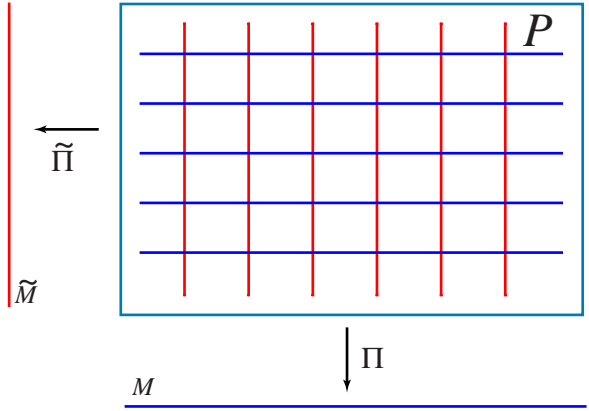}
    \begin{quote}
	\caption[xxx]{A bifibration where the vertical fibers are diffeomorphic 
	to a cover of $\Mtil$ and the horizontal ones are diffeomorphic to a 
	cover of $M$.}
	\label{fig:bifibration}
    \end{quote}
\end{figure}

We introduce the following terminology illustrated in 
Figure~\ref{fig:bifibration}.  
At a point $p\in P$ we have a splitting 
of the tangent space $T_{p}P = H_{p}\oplus V_{p}$ where the 
``horizontal tangent space'' $H_{p}$ is tangent to the fiber of 
$\Pitil$, and the ``vertical tangent space'' $V_{p}$ is tangent to the 
fiber of $\Pi$.  A symplectic form $\beta$ is said to be 
\emph{bifibration compatible} if for every $p\in P$ one has the 
following nondegeneracy conditions:
\begin{enumerate}
    \item  Given $Y\in V_{p}$, if for all $X\in H_{p}$ one has 
    $\beta(X,Y)=0$ then $Y=0$.

    \item  Given $X\in H_{p}$, if for all $Y\in V_{p}$ one has 
    $\beta(X,Y)=0$ then $X=0$.
\end{enumerate}
This is the coordinate independent way of stating that the matrix
$m_{ij}$ is invertible.  The reader can verify that the symplectic
form given in footnote~\ref{foot:symplectic} fails the above.  Our
abstract scenario is a bifibration\footnote{It may be possible to
generalize from a bifibration to a bifoliation.  It is not clear to
the author what is the most general formulation.} with a bifibration
compatible symplectic form.  We will refer to such a manifold as a
\emph{special symplectic bifibration}.  We believe that the
formulation of canonical transformations in path space in terms of
$\beta$ is probably more fundamental than the use of a generating
function.  This is probably analogous to the ascendant role the
symplectic $2$-form has taken in symplectic geometry because of global
issues.  The $2$-form $\beta$ may play a role in the quantum aspects
of duality maybe in some geometric quantization type of framework.

A scenario for how a natural generating function of type 
(\ref{eq:defF}) might arise is the following.  For any $M$, the 
cotangent bundle $T^{*}M$ is a symplectic manifold.  Firstly, one has 
to investigate whether the cotangent bundle admits a second fibration 
transverse to the defining one.  Secondly, it may be necessary to 
deform the original symplectic structure.  In the case of a Lie group 
$G$, the cotangent bundle is trivial and is thus a product $T^{*}G = 
G\times\mathfrak{g}$ where we have used the metric on $G$ to identify 
the Lie algebra $\mathfrak{g}$ with its vector space dual 
$\mathfrak{g}^{*}$.

\section{Hamiltonian structure}
\label{sec:hamiltonian}

The discussion in the Section~\ref{sec:symplectic} is general and 
makes no reference to the hamiltonian. The hamiltonian only played an 
indirect role because we chose a class of canonical transformations 
which are linear with respect to $dx/d\sigma$ and $\pi(\sigma)$ in 
anticipation of future application to the nonlinear sigma model. The 
nonlinear sigma model has target space a riemannian manifold $M$ with 
metric $g$ and a $2$-form field $B$. The hamiltonian density and the 
momentum density are respectively given by
\begin{eqnarray}
    \mathcal{H} & = & \half g^{ij}(x) 
    \left(\pi_{i} - B_{ik}\frac{dx^{k}}{d\sigma}\right)
     \left(\pi_{j} - B_{jl}\frac{dx^{l}}{d\sigma}\right)
   + \half g_{ij}(x) \frac{dx^{i}}{d\sigma}\frac{dx^{j}}{d\sigma}\;,
    \label{eq:hamiltonian}  \\
    \mathcal{P} & = & \pi_{i}(\sigma)\,\frac{dx^{i}}{d\sigma}\;.
    \label{eq:momentum}
\end{eqnarray}
We are interested whether we can find a canonical transformation with 
generating function of type (\ref{eq:defF}) which will map the 
hamiltonian density and momentum density into that of another sigma 
model (the dual sigma model) characterized by target space $\Mtil$, 
metric tensor $\gtil$ and $2$-form $\Btil$.

It winds up that working in coordinates is not the best way of 
attacking the problem.  It is best to use moving frames \`a la Cartan 
and Chern.  Let $(\theta^{1},\ldots,\theta^{n})$ be a local 
orthonormal coframe\footnote{Because we will be working in orthonormal 
frames we do not distinguish an upper index from a lower index in a 
tensor.} for $M$.  The Cartan structural equations are
\begin{eqnarray*}
    d\theta^{i} & = & -\omega_{ij}\wedge\theta^{j}\;, \\
    d\omega_{ij} & = & -\omega_{ik}\wedge\omega_{kj}
    +\half R_{ijkl}\theta^{k}\wedge\theta^{l}\;,
\end{eqnarray*}
where $\omega_{ij}=-\omega_{ji}$ is the riemannian 
connection\footnote{The riemannian connection is the unique torsion 
free metric compatible connection.  A metric compatible connection 
will also be referred to as an orthogonal connection.  In general an 
orthogonal connection can have torsion.}.  Next we define $dx/d\sigma$ 
in the orthonormal frame to be $\xs{\relax}$ by requiring that 
$\theta^{i}=\xs{i}d\sigma$.  If $\pi$ is now the canonical momentum 
density in the orthonormal frame then in this frame 
(\ref{eq:hamiltonian}) and (\ref{eq:momentum}) become
\begin{eqnarray}
    \mathcal{H} & = & \half (\pi_{i}-B_{ik}\xs{k})
    (\pi_{i}-B_{il}\xs{l}) + \half \xs{i}\xs{i}\;,
    \label{eq:hamframe}  \\
    \mathcal{P} & = & \pi_{i}\xs{i}
    = (\pi_{i}-B_{ij}\xs{j})\xs{i}\;.
    \label{eq:momframe}
\end{eqnarray}
In this coframe we can write (\ref{eq:beta}) as
\begin{equation}
    \beta = -\half l_{ij}(x,\xtil) \theta^{i}\wedge \theta^{j} + 
    m_{ij}(x,\xtil) \thetatil^{i}\wedge \theta^{j} + \half 
    \ltil_{ij}(x,\xtil) \thetatil^{i}\wedge \thetatil^{j}\;.    
    \label{eq:betaframe}
\end{equation}
We use the same letters $l,m,\ltil$ but the meaning above is different 
from (\ref{eq:beta}).  In this notation equations (\ref{eq:dualpi}) 
and (\ref{eq:dualpitil}) become
\begin{eqnarray}
    \pi_{i}(\sigma) & = & m_{ji}(x,\xtil)\xstil{j}
    	+l_{ij}(x,\xtil)\xs{j}\;,
    \label{eq:dualpiframe}  \\
    \pitil_{i}(\sigma) & = & m_{ij}(x,\xtil)\xs{j}
    	+ \ltil_{ij}(x,\xtil)\xstil{j}\;,
    \label{eq:dualpitilframe}
\end{eqnarray}
In matrix notation the above may be written as
$$
    \left(
    \begin{array}{cc}
        m^{t} & 0  \\
        -\ltil & I
    \end{array}
    \right)
    \left(
    \begin{array}{c}
        \xstil{\relax}  \\
        \pitil
    \end{array}
    \right)
    =
    \left(
    \begin{array}{cc}
        -l & I  \\
        m & 0
    \end{array}
    \right)
    \left(
    \begin{array}{c}
        \xs{\relax}  \\
        \pi
    \end{array}
    \right)
$$
Rewrite the above in the form
$$
    \left(
    \begin{array}{cc}
        m^{t} & 0  \\
        -\ntil & I
    \end{array}
    \right)
    \left(
    \begin{array}{c}
        \xstil{\relax}  \\
        \pitil - \Btil\xstil{\relax}
    \end{array}
    \right)
    =
    \left(
    \begin{array}{cc}
        -n & I  \\
        m & 0
    \end{array}
    \right)
    \left(
    \begin{array}{c}
        \xs{\relax}  \\
        \pi- B\xs{\relax}
    \end{array}
    \right)\;,
$$
where
\begin{eqnarray}
    n & = & l - B\;,
    \label{eq:defn}  \\
    \ntil & = & \ltil - \Btil\;.
    \label{eq:defntil}
\end{eqnarray}
The rewriting above is closely related to (\ref{eq:Y-matrix}), see 
below.
This equation is not very interesting in this form but it becomes much 
more interesting when rewritten as
\begin{eqnarray}
    \left(
    \begin{array}{c}
        \xstil{\relax}  \\
        \pitil - \Btil\xstil{\relax}
    \end{array}
    \right)
    &=&
    \left(
    \begin{array}{cc}
        m^{t} & 0  \\
        -\ntil & I
    \end{array}
    \right)^{-1}
    \left(
    \begin{array}{cc}
        -n & I  \\
        m & 0
    \end{array}
    \right)
    \left(
    \begin{array}{c}
        \xs{\relax}  \\
        \pi - B\xs{\relax}
    \end{array}
    \right)
    \nonumber\\
    &=&
    \left(
    \begin{array}{cc}
        -\mtinv n & \mtinv  \\
        -\ntil\mtinv n + m & \ntil \mtinv
    \end{array}
    \right)
    \left(
    \begin{array}{c}
        \xs{\relax}  \\
        \pi - B\xs{\relax}
    \end{array}
    \right)\;.
    \label{eq:canmatrix}
\end{eqnarray}
Notice that equation (\ref{eq:canmatrix}) gives us a linear
transformation between $(\xs{\relax},\pi- B\xs{\relax})$ and
$(\xstil{\relax},\pitil-\Btil\xstil{\relax})$.  The preservation of
the hamiltonian density means that this linear transformation must be
in $\Orth(2n)$.  If in addition you want to preserve the momentum
density then this transformation must be in $\OrthQ(n,n)$, the
group of $2n\times 2n$ matrices isomorphic to $\Orth(n,n)$ which
preserves the quadratic form
\begin{equation}
    Q = \left(
    \begin{array}{cc}
        0 & I_{n}  \\
        I_{n} & 0
    \end{array}
    \right)\;.
    \label{eq:defQ}
\end{equation}
In the formula above, $I_{n}$ is the $n\times n$ identity matrix. 
Properties of $\OrthQ(n,n)$ and its relation with $\Orth(2n)$ are
reviewed in Appendix~\ref{sec:orthogonal}.  They key
observation\footnote{I do not understand geometrically why $\beta$
automatically induces this pseudo-orthogonal matrix.} is that the
matrix appearing in (\ref{eq:canmatrix}) is automatically in
$\OrthQ(n,n)$ which means that our canonical transformation
automatically preserves the canonical momentum density
(\ref{eq:momframe}).  As previously mentioned to preserve the
hamiltonian density (\ref{eq:hamframe}) is it necessary that the
matrix above also be in $\Orth(2n)$.  Thus the matrix
\begin{equation}
    \left(
    \begin{array}{cc}
        -\mtinv n & \mtinv  \\
        -\ntil\mtinv n + m & \ntil \mtinv
    \end{array}
    \right)
    \label{eq:matrix}
\end{equation}
must be in $\Orth(2n)\cap\OrthQ(n,n)$, a compact group locally 
isomorphic to $\Orth(n)\times\Orth(n)$, see 
Appendix~\ref{sec:orthogonal}.  Using the equations in the appendix we 
learn that the condition that (\ref{eq:matrix}) be in the intersection 
$\Orth(2n)\cap\OrthQ(n,n)$ is that
\begin{eqnarray}
    mm^{t} & = & I - \ntil^{2}\;,
    \label{eq:mmt}  \\
    m^{t}m & = & I- n^{2}\;,
    \label{eq:mtm}  \\
    -mn & = & \ntil m\;.
    \label{eq:mn}
\end{eqnarray}
We can now simplify (\ref{eq:matrix}) to
\begin{equation}
    \left(
    \begin{array}{cc}
        -\mtinv n & \mtinv  \\
        \mtinv & -\mtinv n
    \end{array}
    \right)
    \label{eq:matrix1}
\end{equation}

To better understand the above is is worthwhile using the conjugation 
operation (\ref{eq:conjugation}) and switch the quadratic 
from from $Q$ to
$$
    \left(
    \begin{array}{cc}
        -I & 0  \\
        0 & +I
    \end{array}
    \right)\;.
$$
Under this conjugation operation (\ref{eq:canmatrix}) becomes
\begin{eqnarray}
   & & \left(
    \begin{array}{c}
        \xstil{\relax} -(\pitil - \Btil\xstil{\relax}) \\
        \xstil{\relax} +(\pitil - \Btil\xstil{\relax})
    \end{array}
    \right)
    \nonumber\\
    &&\qquad =
    \left(
    \begin{array}{cc}
        -\mtinv(I+n) & 0  \\
        0 &  \mtinv (I-n)
    \end{array}
    \right)
    \left(
    \begin{array}{c}
        \xs{\relax}  -(\pi - B\xs{\relax})\\
        \xs{\relax} +(\pi - B\xs{\relax})
    \end{array}
    \right)\;.
    \label{eq:canmatrix1}
\end{eqnarray}
This leads to the pair of equations
\begin{eqnarray}
    (\pitil - \Btil\xstil{\relax}) +\xstil{\relax} & = & 
    +T_{+}\left[(\pi - B\xs{\relax}) +\xs{\relax}\right]\;,
    \label{eq:canplus}  \\
    (\pitil - \Btil\xstil{\relax}) -\xstil{\relax} & = & 
    -T_{-}\left[(\pi - B\xs{\relax}) -\xs{\relax}\right]\;,
    \label{eq:canminus}
\end{eqnarray}
where
\begin{equation}
    T_{\pm}=  \mtinv (I \mp n) \in \Orth(n)\;.
    \label{eq:defTpm}
\end{equation}
An equivalent way of writing the above is $m=T_{\pm}(I \pm n)$.  Also 
note that $T_{+}$ and $T_{-}$ are not independent.  They are related 
by $T^{-1}_{-} T_{+}= (I+n)^{-1}(I-n)$ which is the Cayley transform 
of $n$.  It is often convenient to think that (\ref{eq:betaframe}) is 
determined by two orthogonal matrices $T_{\pm}\in\Orth(n)$ with
\begin{equation}
    n = -(T_{+}+T_{-})^{-1}(T_{+}-T_{-}).
    \label{eq:solnn}
\end{equation}

\section{Gauge invariance}
\label{sec:gauge-invariance}

It is well known that the sigma model $(M,g,B)$ has a gauge invariance 
given by $B \to B + dA$ where $A$ is a $1$-form on $M$.  We can 
manifest these gauge transformations within the class 
(\ref{eq:defTpm}) of canonical transformation by considering $\int 
\alpha \to \int(\alpha+A)$ which transforms $\pi$ appropriately. An 
observation and a change of viewpoint will give us a manifestly gauge 
invariant formulation. Notice that both the left hand side and right 
hand side of equation (\ref{eq:canmatrix1}) is manifestly gauge 
invariant. This suggests that $m,n,\ntil$ may be gauge invariant. 
Looking at (\ref{eq:defn}) and (\ref{eq:defntil}) and incorporating 
the remark about how we implement gauge invariance we see that $n$ 
and $\ntil$ are gauge invariant quantities, \emph{i.e.}, the gauge 
transformations are implemented by shifting $l,\ltil$ respectively 
by $dA$ and $d\widetilde{A}$. This suggest that instead of working 
with $\beta$ it may be worthwhile to work with $\gamma$ defined by
\begin{equation}
    \gamma = -\half n_{ij}(x,\xtil) \theta^{i}\wedge \theta^{j} + 
    m_{ij}(x,\xtil) \thetatil^{i}\wedge \theta^{j} + \half 
    \ntil_{ij}(x,\xtil) \thetatil^{i}\wedge \thetatil^{j}    
    \label{eq:gammaframe}
\end{equation}
where $\gamma$ is not closed but satisfies
\begin{equation}
    d\gamma = H - \Htil
    \label{eq:dgamma}
\end{equation}
where $H=dB$ and $\Htil=d\Btil$. More correctly one has $d\gamma = 
\Pi^{*}H -\Pitil^{*}\Htil$. We have now achieved a gauge invariant 
formulation.

\section{The geometry of $P$}
\label{sec:geometry}

To gain further insight into relations between the geometry of $M$ 
and $\Mtil$ is it best to work in $P$ which you may think of it 
locally being $M\times\Mtil$. We can use the freedom of working in 
$P$ to simplify results and then project back to either $M$ or $\Mtil$.

There are two closely related ways of simplifying the geometry.  One 
way is to work in the bundle of orthonormal frames.  The other is to 
\emph{adapt} the orthonormal frames to the problem at hand similar to 
the way one uses Darboux frames to study surfaces in classical 
differential geometry.  The former gives a global formulation but the 
latter is more familiar to physicists hence we choose the latter.
All our computations will be local and can be patched together 
to define global objects.

The first thing to observe is that the existence of the double
fibration allows us to naturally define a riemannian metric on $P$ by
pulling back the metrics on $M$ and $\Mtil$ and declaring that the
fibers are orthogonal to each other.  In a similar fashion we pullback
local coframes and get local coframes on $P$.  These orthonormal
coframes satisfy the Cartan structural equations
\begin{eqnarray}
    d\theta^{i} & = & -\omega_{ij}\wedge\theta^{j}\;,
    \label{eq:cartan1} \\
    d\thetatil^{i} & = & -\omegatil_{ij}\wedge\thetatil^{j}\;,
    \label{eq:cartan1til} \\
    d\omega_{ij} & = & -\omega_{ik}\wedge\omega_{kj}
    +\half R_{ijkl}\theta^{k}\wedge\theta^{l}\;,
    \label{eq:cartan2} \\
    d\omegatil_{ij} & = & -\omegatil_{ik}\wedge\omegatil_{kj}
    +\half \Rtil_{ijkl}\thetatil^{k}\wedge\thetatil^{l}\;.
    \label{eq:cartan2til}
\end{eqnarray}
Once we begin working on $P$ then we have the freedom to independently 
rotate $\theta$ and $\thetatil$ at each point.  Once we do this these 
coframes will no longer be pullbacks but this doesn't matter because 
it does not change the metric on each fiber.  We are going to exploit 
this freedom to relate the geometry of $M$ to that of $\Mtil$ in a way 
similar to the way the intrinsic curvature of a submanifold is 
related to the total curvature of the space and the curvature of the 
normal bundle.  Note that with these choices there is a natural group 
of $\Orth(n)\times\Orth(n)$ gauge transformations on the tangent bundle of 
$P$ which is compatible with the metric structure and the 
bifibration.

\section{Constraints from the algebraic structure of $\gamma$}
\label{sec:algebraicconstraints}

First we derive various constraints that follow from the algebraic 
constraints on $\gamma$ imposed by the preservation of $\mathcal{H}$ 
and $\mathcal{P}$.  Equations~(\ref{eq:mtm}) and (\ref{eq:mn}) tell us 
that 
\begin{equation}
    m=T(I+n)\quad\mbox{and}\quad \ntil = -TnT^{t}
    \label{eq:mreduction}
\end{equation}
where $T\in\Orth(n)$.  Since $T$ ``connects'' a $\theta$ to a 
$\thetatil$ we see that its covariant differential is given by
\begin{equation}
    dT_{ij}+\omegatil_{ik}T_{kj}+\omega_{jk}T_{ik}= 
    +T_{ijk}\theta^{k} - \Ttil_{ijk}\thetatil^{k}\;,
    \label{eq:covderT}
\end{equation}
where the components of the covariant differential in the $M$ 
direction is $+T_{ijk}$ and in the $\Mtil$ direction is $-\Ttil_{ijk}$.  
The negative sign is introduced for future convenience.  Notice that 
$T_{ijk}$ and $\Ttil_{ijk}$ are tensors defined on $P$ whose existence 
is guaranteed by the existence of the tensor $T_{ij}$ on $P$.

We now invoke a ``symmetry breaking mechanism'' to reduce the
structure group of gauge transformations from $\Orth(n)\times\Orth(n)$
to $\Orth(n)$.  At each point in $P$ we can rotate $\thetatil$ (or
$\theta$) and make $T=I$ because under these gauge transformations $T
\to \Rtil T R^{-1}$ where $(R,\Rtil)\in \Orth(n)\times\Orth(n)$.  The
isotropy group of $T=I$ is the diagonal $\Orth(n)$.  This is no
different than giving a scalar field a vacuum expectation value to
break the symmetry.  This symmetry breaking leads to an identification
at each point of $P$ of the ``vertical'' and ``horizontal'' tangent
spaces.  This does not tell us that the metrics are the same but
allows us to identify an orthonormal frame in one with an orthonormal
frame in the other.  Let us be a bit more precise and abstract on the
reduction of the structure group and the identification of the
``vertical'' and ``horizontal'' tangent spaces.  We already mentioned
that at $p\in P$ one has $T_{p}P = H_{p}\oplus V_{p}$.  The tensor
$m(p)$ may be viewed as an element of $V^{*}_{p}\otimes H^{*}_{p}$. 
Because there is a metric on $V_{p}$ we can reinterpret $m$ as giving
us an invertible linear transformation $\check{m}:H_{p}\to V_{p}$.  We
also have a metric on $H_{p}$ and thus we can study the orbit of
$m(p)$ under the action of $\Orth(n)\times\Orth(n)$.  Our previous
discussion shows that a ``canonical'' form for $m(p)$ may be taken to
be $m(p)=I+n(p)$ with isotropy group being the diagonal $\Orth(n)$. 
If $(e_{1},\ldots,e_{n})$ is an orthonormal basis at $H_{p}$ and
$(\etil_{1},\ldots,\etil_{n})$ is the corresponding orthonormal basis
at $V_{p}$ then they are related by $\check{m}(p)e_{i}
=\etil_{j}(\delta_{ji}+n_{ji}(p))$.

From now on we assume we have adapted our coframes such that $T=I$ and
\begin{eqnarray}
    m_{ij} & = & \delta_{ij} + n_{ij}\,,
    \label{eq:mreduced}  \\
    n_{ij} & = & \ntil_{ij}\;.
    \label{eq:nntilreduced}
\end{eqnarray}
In this frame, $\gamma$ simplifies to
\begin{equation}
    \gamma = \thetatil^{i}\wedge\theta^{i} + 
    n_{ij}\thetatil^{i}\wedge\theta^{j}
    -\half n_{ij}\theta^{i}\wedge\theta^{j}
    -\half n_{ij}\thetatil^{i}\wedge\thetatil^{j}\;.
    \label{eq:gammareduced}
\end{equation}
The duality equations are particularly simple now and they 
are given by
\begin{eqnarray}
    (\pitil - \Btil\xstil{\relax}) +\xstil{\relax} & = & 
    (\pi - B\xs{\relax}) +\xs{\relax}\;,
    \label{eq:canplus1}  \\
    (\pitil - \Btil\xstil{\relax}) -\xstil{\relax} & = & 
    -T_{-}\left[(\pi - B\xs{\relax}) -\xs{\relax}\right]\;,
    \label{eq:canminus1}
\end{eqnarray}
Where the orthogonal matrix $T_{-}$ is the Cayley transform of $n$:
\begin{equation}
    T_{-}= \frac{I+n}{I-n}\;.
    \label{eq:TCayley}
\end{equation}
The matrix $T_{-}$ is not arbitrary because there are constraints on 
$n_{ij}$ as we will see later on. Without constraints on $T_{-}$ there 
are interesting solutions to (\ref{eq:canplus1}) and 
(\ref{eq:canminus1}) which map spaces of constant positive curvature 
into spaces of negative constant curvature or more generally dual 
symmetric spaces\footnote{O. Alvarez, unpublished.}.

We can now exploit equation~(\ref{eq:covderT}) to relate the 
connections in the adapted coframing. Inserting $T=I$ into the above 
leads to
\begin{equation}
    \omegatil_{ij}-\omega_{ij} = 
    +T_{ijk}\theta^{k}-\Ttil_{ijk}\thetatil^{k}\;.
    \label{eq:reducedconnection}
\end{equation}
Thus we see that in the reduction of the structure group we have 
generated torsion and that this torsion satisfies $T_{ijk}=-T_{jik}$ 
and $\Ttil_{ijk}=-\Ttil_{jik}$.
We now define an orthogonal connection on our adapted frames by
\begin{equation}
    \psi_{ij}=\omega_{ij} + T_{ijk}\theta^{k}
    =\omegatil_{ij}+\Ttil_{ijk}\thetatil^{k}\;.
    \label{eq:psiconnection}
\end{equation}
First we define the components of the covariant 
derivatives of $T$ and 
$\Ttil$ by
\begin{eqnarray}
    dT_{ijk} + (\omega\cdot T)_{ijk} & = & T'_{ijkl}\theta^{l} + 
    T''_{ijkl}\thetatil^{l}\;,
    \label{eq:Tder}  \\
    d\Ttil_{ijk} + (\omegatil\cdot \Ttil)_{ijk} & = & \Ttil'_{ijkl}\theta^{l} + 
    \Ttil''_{ijkl}\thetatil^{l}\;.
    \label{eq:Ttilder}
\end{eqnarray}
In the above $(\omega\cdot T)$ and $(\omegatil\cdot \Ttil)$ are 
abbreviations for standard expressions.  We have chosen to use the 
connections $\omega$ and $\omegatil$ rather than $\psi$ in the 
definition of the covariant derivative for the following reasons: if 
$T_{ijk}$ is the pullback of a tensor on $M$ then $T''_{ijkl}=0$; if 
$\Ttil_{ijk}$ is the pullback of a tensor on $\Mtil$ then 
$\Ttil'_{ijkl}=0$. A notational remark is that a primed tensor denoted the 
covariant derivative in the $M$ direction and a doubly primed tensor 
denotes the covariant derivative in the $\Mtil$ direction. Doubly 
primed does not mean second derivative.

The curvature of this connection may be computed by either using the 
expression involving $\omega$ or the one involving $\omegatil$.  A 
straightforward computation of the curvature matrix $2$-form
\begin{equation}
    \Psi_{ij} = d\psi_{ij}  + \psi_{ik}\wedge\psi_{kj}
    \label{eq:Psi}
\end{equation}
in these two ways leads to the following expressions
\begin{eqnarray*}
    \Psi_{ij} &=& -T''_{ijlm} 
    \theta^{l}\wedge\thetatil^{m}  \\
     & + & \half\left[ R_{ijlm} - (T'_{ijlm}-T'_{ijml})
     +(T_{ikl}T_{kjm}-T_{ikm}T_{kjl})\right]
     \theta^{l}\wedge\theta^{m}\;,
\end{eqnarray*}
and
\begin{eqnarray*}
    \Psi_{ij} & = & -\Ttil'_{ijlm} 
    \thetatil^{l}\wedge\theta^{m}  \\
     & + & \half\left[ \Rtil_{ijlm} - (\Ttil''_{ijlm}-\Ttil''_{ijml})
     +(\Ttil_{ikl}\Ttil_{kjm}-\Ttil_{ikm}\Ttil_{kjl})\right]
     \thetatil^{l}\wedge\thetatil^{m}\;.
\end{eqnarray*}
Comparing these two expression we learn that the curvature two form 
matrix is given by
\begin{equation}
    \Psi_{ij} = d\psi_{ij}  + \psi_{ik}\wedge\psi_{kj} = -T''_{ijlm} 
    \theta^{l}\wedge\thetatil^{m}\;.
    \label{eq:psicurvature}
\end{equation}
The following constraints must also hold
\begin{eqnarray}
    R_{ijlm} - (T'_{ijlm}-T'_{ijml})
     +(T_{ikl}T_{kjm}-T_{ikm}T_{kjl}) & = & 0 \;,
    \label{eq:Rparallel}  \\
    \Rtil_{ijlm} - (\Ttil''_{ijlm}-\Ttil''_{ijml})
     +(\Ttil_{ikl}\Ttil_{kjm}-\Ttil_{ikm}\Ttil_{kjl}) & = & 0 \;,
    \label{eq:Rtilparallel}  \\
    T''_{ijlm} + \Ttil'_{ijml} & = & 0 \;.
    \label{eq:Tparallel}
\end{eqnarray}
Form (\ref{eq:psicurvature}) is reminiscent of a K\"{a}hler manifold 
where the curvature is of type $dz\wedge d\bar{z}$ and there are no 
$dz\wedge dz$ or $d\bar{z}\wedge d\bar{z}$ components. The absence of 
these many curvature components is due to the reduction of the 
structure group from $\Orth(2n)$ to $\Orth(n)$ at the expense of 
generating torsion.

There are a variety of equivalent ways of interpreting the above.  The 
most geometric is to observe that $\psi_{ij}$ defines a connection on 
$P$ and thus a connection when restricted to any of the fibers.  For 
example, let $M_{\xtil} = \Pitil^{-1}(\xtil)$ be a horizontal fiber.  
Notice that along this fiber $\thetatil=0$ and thus $\Psi_{ij}=0$.  
Since $M_{\xtil}$ is isometric to $M$ we have found a flat orthogonal 
connection (generally with torsion) on $M$. Note that 
this is true for all horizontal fibers. One can make a similar 
statement about the vertical fibers. We have our first major result.
\begin{quote}
    \em 
    Target space duality requires that the manifolds $M$ and $\Mtil$ 
    respectively admit flat orthogonal connections.  The connection 
    $\psi_{ij}$ is flat when restricted to either $M$ or $\Mtil$.
    \rm
\end{quote}
At a more algebraic level equations~(\ref{eq:Rparallel}) and
(\ref{eq:Rtilparallel}) are the standard equations for
``parallelizing'' the curvature by torsion.  A manifold $M$ is said to
be parallelizable if the tangent bundle is a product bundle $TM =
M\times\bbR^{n}$.  This means that you can globally choose a frame on
$M$.  The existence of a flat connection on a manifold does not imply
parallelizability.  The reason is that in a non-simply connected
manifold there is an obstruction to globally choosing a frame if there
is holonomy.  If the manifold is simply connected and the connection
is flat then it is parallelizable.  Finally we observe that if a
manifold is parallelizable then there are an infinite number of other
possible parallelizations\footnote{I would like to thank I.M. Singer
for the ensuing argument.}.  Assume we have an orthogonal
parallelization, \emph{i.e.}, a choice of orthonormal frame at each
point.  Given any other orthogonal parallelization we can always make
a rotation point by point so that both frames agree at the point. 
Thus the space of all orthogonal parallelizations is given by the set
of maps from $M$ to $\Orth(n)$.

Note that given two distinct points $\xtil_{1},\xtil_{2}\in\Mtil$, the 
tensor $T_{ijk}$ on the respective horizontal fibers $M_{\xtil_{1}}$ 
and $M_{\xtil_{2}}$ do not have to be the same.  There are many flat 
orthogonal connections on $M$ as can be seen by a variant 
parallelizability argument.  In fact you could in principle have a 
multiparameter family parametrized by $\Mtil$.

There is a special case of interest when $T_{ijk}$ is the pullback of
a tensor on $M$.  In this case a previous remark tells us that
$T''_{ijkl}=0$ and consequently by (\ref{eq:Tparallel}) we also have
$\Ttil'_{ijkl}=0$.  Therefore $\Ttil_{ijk}$ is also the pullback of a
tensor on $\Mtil$.  This means that the same torsion tensors make the
connection flat on all the fibers.  Note that in this case
$\Psi_{ij}=0$ and the orthogonal connection $\psi_{ij}$ is a flat
connection on $P$.
\begin{quote}
    \em
    If $T_{ijk}$ is the pullback of a tensor on $M$ then $\Ttil_{ijk}$ is 
    the pullback of a tensor on $\Mtil$ and $\Psi_{ij}=0$. In this 
    case  $\psi_{ij}$ is a flat connection on $P$.
    \rm
\end{quote}

\section{Simple examples}
\label{sec:dgammaconstraints}

The equation $d\gamma = H - \Htil$ introduces relations among 
$H,\Htil,T_{ijk}$ and $\Ttil_{ijk}$. First we point out some facts.

\subsection{The case of $n_{ij}=0$}
\label{sec:nequals0}

As a warmup we study the case where $n_{ij}=0$. In this case 
$\gamma = \thetatil^{i}\wedge\theta^{i}$ and we compute $d\gamma$ by 
using the Cartan structural equations (\ref{eq:cartan1}),
(\ref{eq:cartan1til}) and the condition which follows from the 
reduction of the symmetry group (\ref{eq:reducedconnection}). A brief 
computation yields
$$
    d\gamma = T_{kij}\theta^{i}\wedge\theta^{j}\wedge\thetatil^{k}
    -\Ttil_{ijk}\theta^{i}\wedge\thetatil^{j}\wedge\thetatil^{k}\;.
$$
First we learn that the $3$-forms $H$ and $\Htil$ vanish.  Next we see 
that $T_{kij}=T_{kji}$ and $\Ttil_{ijk}=\Ttil_{ikj}$.  We remind the 
reader that a tensor $S_{ijk}$ which is skew symmetric under $i\leftrightarrow 
j$ and symmetric under $j\leftrightarrow k$ is zero.  Thus we conclude 
that $T_{ijk}=\Ttil_{ijk}=0$.  It follows from equations 
(\ref{eq:Rparallel}) and (\ref{eq:Rtilparallel}) that 
$R_{ijkl}=\Rtil_{ijkl}=0$.  Since the Riemannian curvatures vanish we 
know that $M$ and $\Mtil$ are manifolds with universal cover 
$\bbR^{n}$.  There are no other possibilities if $n_{ij}=0$.  For 
example you can have $M=\mathbb{T}^{k}\times\bbR^{n-k}$.  This is the 
case of abelian duality.  Other potential singular cases of interest 
are orbifolds or cones which are flat but have holonomy due to the 
presence of singularities.

\subsection{The case of a Lie group}
\label{sec:Liegroup}

We verify that the standard nonabelian duality results are 
reproducible in this formalism.  We present a schematic discussion 
here because the Lie group example is a special case of a more general 
result presented in Section~\oaref{sec:cotangent}{2.2.1}.
Let $G$ a compact simple Lie group 
with Lie algebra $\mathfrak{g}$.  Let $(e_{i},\ldots,e_{n})$ is an 
orthonormal basis for $\mathfrak{g}$ with respect to the Killing form.  
The structure constants $f_{ijk}$ are defined by 
$[e_{i},e_{j}]=f_{kij} e_{k}$.  In this case the structure 
constants are totally antisymmetric.  Let $\theta^{i}$ be the 
associated Maurer-Cartan forms satisfying the Maurer-Cartan equations
\begin{equation}
    d\theta^{i}= -\half f_{ijk}\theta^{j}\wedge\theta^{k}\;.
    \label{eq:Maurer-Cartan}
\end{equation}
Because of the Killing form we can identify the Lie algebra
$\mathfrak{g}$ with its vector space dual $\mathfrak{g}^{*}$.  We
choose $P$ to be the cotangent bundle $T^{*}G$ which is a product
bundle $T^{*}G = G\times\mathfrak{g}^{*} = G\times\mathfrak{g}$.  If
$(p_{1},\ldots,p_{n})$ are the standard coordinates on the cotangent
bundle with respect to the orthonormal frame then the we take $\alpha$
in (\ref{eq:defF}) to be $\alpha = p_{i}\theta^{i}$, the canonical
$1$-form on $T^{*}G$.  Therefore $\beta=d\alpha$ is the standard
symplectic form on $T^{*}G$ given by
\begin{equation}
    \beta = dp_{i}\wedge\theta^{i} - \half 
    p_{i}f_{ijk}\theta^{j}\wedge\theta^{k}\;.
    \label{eq:Gsymplectic}
\end{equation}
By looking at reference \cite{Alvarez:1995uc} one can see that the
orthonormal coframe $(\thetatil^{1},\ldots,\thetatil^{n})$ on the
fiber $\mathfrak{g}^{*}$ is given by $dp_{j}
=\thetatil^{i}(\delta_{ij} + f_{kij}p_{k})$.  This suggests that
$m_{ij}=(\delta_{ij} + f_{kij}p_{k})$ and that in this basis the
symmetry breaking is manifest and thus $n_{ij}=f_{kij}p_{k}$.  Thus we
expect that $\gamma$ is given by
\begin{equation}
    \gamma = -\half f_{kij}p_{k} \theta^{i}\wedge \theta^{j} + 
    (\delta_{ij}+f_{kij}p_{k}) \thetatil^{i}\wedge \theta^{j} 
    -\half f_{kij}p_{k} \thetatil^{i}\wedge \thetatil^{j}\;. 
    \label{eq:G-gamma}
\end{equation}
Note that $d\gamma = -\Htil$ because the 
modification of going from the closed form $\beta$ to $\gamma$ 
involved a term of the type $n_{ij}\thetatil^{i}\wedge\thetatil^{j}$. 
To verify this we observe that $\thetatil^{i}=dp_{j}m^{-1}_{ji}$ and 
thus $n_{ij}\thetatil^{i}\wedge\thetatil^{j}$ only depends on $p$ and 
$dp$, therefore, its exterior derivative can only be of type 
$dp\wedge dp\wedge dp \sim \thetatil\wedge\thetatil\wedge\thetatil$. 
In fact $\half f_{kij}p_{k} \thetatil^{i}\wedge \thetatil^{j}$ is the 
standard representation for the $2$-form $\Btil$.

If we write $d\thetatil^{i}=-\half 
\ftil_{ijk}\theta^{j}\wedge\theta^{k}$ then a straightforward exercise 
shows that
$$
	\ftil_{ijk}=(m_{jm}f_{mkl}-m_{km}f_{mjl})m^{-1}_{li}\;.
$$
By using (\ref{eq:rotationcoeff}) one can compute $\omegatil_{ij}$.  
It is now an algebraic exercise to compute parallelizing torsions 
$T_{ijk}$ and $\Ttil_{ijk}$.

\section{The case of a general connection $\psi$}

\subsection{General theory}
\label{sec:psi}

We already saw that the connection $\psi_{ij}$ on $P$ gives a flat 
connection on both $M$ and $\Mtil$, a necessary condition for $M$ and 
$\Mtil$ to be target space duals of each other.  We are going to take 
the following approach.  Assume we are given a $\psi_{ij}$ on $P$, how 
do we determine $n_{ij}$?  We will derive  PDEs that $n_{ij}$ must 
satisfy.  If there exist solutions to these PDEs then we automatically 
have a duality between the sigma model on $M$ and the one on $\Mtil$.  
for It is worthwhile to rewrite the Cartan structural equations in 
terms of $\psi_{ij}$:
\begin{eqnarray}
    d\theta^{i} & = & -\psi_{ij}\wedge\theta^{j} - \half 
    f_{ijk}\theta^{j}\wedge\theta^{k}\;,
    \label{eq:cartan1psi}  \\
    d\thetatil^{i} & = & -\psi_{ij}\wedge\thetatil^{j} - \half 
    \ftil_{ijk}\thetatil^{j}\wedge\thetatil^{k}\;,
    \label{eq:cartan1psitil} \\
    d\psi_{ij} & = & -\psi_{ik}\wedge\psi_{kj} - 
    T''_{ijlm}\theta^{l}\wedge\thetatil^{m}\;.
    \label{eq:cartan2psi}
\end{eqnarray}
where $f_{ijk}=-f_{ikj}$, $\ftil_{ijk}=-\ftil_{ikj}$ and
$T''_{ijkl}=-T''_{jikl}$.  The structure functions $f_{ijk}$ and
$\ftil_{ijk}$ are related to $T_{ijk}$ and $\Ttil_{ijk}$ by
\begin{eqnarray}
    f_{ijk}=T_{ijk}-T_{ikj}, & \quad & 
    T_{ijk}=\half(f_{ijk}-f_{jik}-f_{kij}) \;,
    \label{eq:ftoT}  \\
    \ftil_{ijk}=\Ttil_{ijk}-\Ttil_{ikj}, & \quad & 
    \Ttil_{ijk}=\half(\ftil_{ijk}-\ftil_{jik}-\ftil_{kij}) \;.
    \label{eq:ftiltoTtil}
\end{eqnarray}
We define the components $n'_{ijk},
n''_{ijk},f'_{ijkl},f''_{ijkl},\ftil'_{ijkl},\ftil''_{ijkl}$ of the
covariant derivatives of $n_{ij},f_{ijk},\ftil_{ijk}$ with respect to
the connection $\psi_{ij}$ by
\begin{eqnarray}
    dn_{ij}+\psi_{ik}n_{kj}+\psi_{jk}n_{ik} &=& n'_{ijk}\theta^{k} + 
    n''_{ijk}\thetatil^{k}\;.
    \label{eq:covdernpsi} \\
    df_{ijk}+\psi_{il}f_{ljk} +\psi_{jl}f_{ilk} +\psi_{kl}f_{ijl} &=&
    f'_{ijkl}\theta^{l}+f''_{ijkl}\thetatil^{l}\;,
    \label{eq:covderf} \\
    d\ftil_{ijk}+\psi_{il}\ftil_{ljk} +\psi_{jl}\ftil_{ilk}
    +\psi_{kl}\ftil_{ijl} &=&
    \ftil'_{ijkl}\theta^{l}+\ftil''_{ijkl}\thetatil^{l}\;.
    \label{eq:covderftil}
\end{eqnarray}
There are several important constraints which follow from 
$d^{2}\theta=d^{2}\thetatil=0$:
\begin{eqnarray}
    \left(-f'_{ijkl}+f_{mjk}f_{iml}\right) 
    \theta^{j}\wedge\theta^{k}\wedge\theta^{l} & = & 0\;,
    \label{eq:fbianchi}  \\
    f''_{ijkl} & = & T''_{ijkl}-T''_{ikjl}\;,
    \label{eq:fppTpp}  \\
    \left(-\ftil''_{ijkl}+\ftil_{mjk}\ftil_{iml}\right) 
    \thetatil^{j}\wedge\thetatil^{k}\wedge\thetatil^{l} & = & 0
    \label{eq:ftilbianchi}  \\
    \ftil'_{ijkl} & = & -(T''_{ijlk}-T''_{iklj})\;.
    \label{eq:ftilppTpp}
\end{eqnarray}
Note that  $T''_{ijkl}=0$ if and only if $f''_{ijkl}=\ftil'_{ijkl}=0$,
\emph{i.e.}, $f_{ijk}$ and $\ftil_{ijk}$ are respectively pullbacks in
accord with a previous remark.  The $d^{2}\psi_{ij}=0$ constraints are
not used in this report and will not be given.

To derive the PDE satisfied by $n_{ij}$ we compute $d\gamma$:
\begin{eqnarray}
    d\gamma &=& H - \widetilde{H} \nonumber \\
  & = & -{\frac{1}{2}} \tensor{n'}{\down{i}\down{j}\down{k}} 
    \tensor{\theta}{\up{i}}\wedge 
     \tensor{\theta}{\up{j}}\wedge \tensor{\theta}{\up{k}} 
     + 
  {\frac{1}{2}} \tensor{f}{\down{i}\down{j}\down{k}} 
    \tensor{n}{\up{i}\down{l}} 
     \tensor{\theta}{\up{j}}\wedge 
      \tensor{\theta}{\up{k}}\wedge \tensor{\theta}{\up{l}} \nonumber \\
    &- & 
    {\frac{1}{2}} \tensor{n''}{\down{i}\down{j}\down{k}} 
    \tensor{\theta}{\up{i}}\wedge 
     \tensor{\theta}{\up{j}}\wedge 
      \tensor{\tilde{\theta}}{\up{k}} - 
  \tensor{n'}{\down{i}\down{j}\down{k}} 
   \tensor{\theta}{\up{i}}\wedge 
    \tensor{\theta}{\up{k}}\wedge 
     \tensor{\tilde{\theta}}{\up{j}} \nonumber \\
    & + &  {\frac{1}{2}} \tensor{f}{\down{i}\down{j}\down{k}} 
    \tensor{\theta}{\up{j}}\wedge 
     \tensor{\theta}{\up{k}}\wedge 
      \tensor{\tilde{\theta}}{\up{i}} - 
  {\frac{1}{2}} \tensor{f}{\down{i}\down{j}\down{k}} 
    \tensor{n}{\up{i}\down{l}} 
     \tensor{\theta}{\up{j}}\wedge 
      \tensor{\theta}{\up{k}}\wedge 
       \tensor{\tilde{\theta}}{\up{l}} \nonumber \\
      & + &  
  \tensor{n''}{\down{i}\down{j}\down{k}} 
   \tensor{\theta}{\up{i}}\wedge 
    \tensor{\tilde{\theta}}{\up{j}}\wedge 
     \tensor{\tilde{\theta}}{\up{k}}   - 
  {\frac{1}{2}} \tensor{n'}{\down{i}\down{j}\down{k}} 
    \tensor{\theta}{\up{k}}\wedge 
     \tensor{\tilde{\theta}}{\up{i}}\wedge 
      \tensor{\tilde{\theta}}{\up{j}} \nonumber \\
      & - & 
  {\frac{1}{2}} \tensor{\tilde{f}}{\down{i}\down{j}\down{k}
     } \tensor{\theta}{\up{i}}\wedge 
     \tensor{\tilde{\theta}}{\up{j}}\wedge 
      \tensor{\tilde{\theta}}{\up{k}} - 
  {\frac{1}{2}} \tensor{\tilde{f}}{\down{i}\down{j}\down{k}
     } \tensor{n}{\up{i}\down{l}} 
     \tensor{\theta}{\up{l}}\wedge 
      \tensor{\tilde{\theta}}{\up{j}}\wedge 
       \tensor{\tilde{\theta}}{\up{k}} \nonumber \\
       & - & 
  {\frac{1}{2}} \tensor{n''}{\down{i}\down{j}\down{k}} 
    \tensor{\tilde{\theta}}{\up{i}}\wedge 
     \tensor{\tilde{\theta}}{\up{j}}\wedge 
      \tensor{\tilde{\theta}}{\up{k}} + 
  {\frac{1}{2}} \tensor{\tilde{f}}{\down{i}\down{j}\down{k}
     } \tensor{n}{\up{i}\down{l}} 
     \tensor{\tilde{\theta}}{\up{j}}\wedge 
      \tensor{\tilde{\theta}}{\up{k}}\wedge 
       \tensor{\tilde{\theta}}{\up{l}} \;.
       \label{eq:dgammaexplicit}
\end{eqnarray}
If we write the closed $3$-forms in components as
\begin{equation}
    H = 
    \frac{1}{3!}H_{ijk}\theta^{i}\wedge\theta^{j}\wedge\theta^{k}\,,
    \quad
    \Htil = \frac{1}{3!}\Htil_{ijk}
    \thetatil^{i}\wedge\thetatil^{j}\wedge\thetatil^{k}\,,    
    \label{eq:3forms}
\end{equation}
where $H_{ijk}$ and $\Htil_{ijk}$ are totally skew symmetric
then we immediately see that 
\begin{eqnarray}
    n'_{ijk}+n'_{jki}+n'_{kij} &=& -H_{ijk}
    +(f_{lij}n_{lk}+f_{ljk}n_{li}+f_{lki}n_{lj}) \;,
    \label{eq:Hnppsi} \\
    n''_{ijk}+n''_{jki}+n''_{kij}&=& +\Htil_{ijk}
    +(\ftil_{lij}n_{lk}+\ftil_{ljk}n_{li}+\ftil_{lki}n_{lj}) \;,
    \label{eq:Htilnpppsi} \\
    (n'_{kij}-n'_{kji}) - n''_{ijk} &=& -(f_{kij}-n_{lk}f_{lij})
    = - m_{kl}f_{lij}\;,
    \label{eq:fnpsi} \\
    -n'_{ijk}+(n''_{kij}-n''_{kji}) &=& 
    +(\ftil_{kij}+n_{lk}\ftil_{lij})
    = \ftil_{lij}m_{lk}\;.
    \label{eq:ftilnpsi}
\end{eqnarray}
The number of linearly independent equations above is 
$\frac{1}{3}n(n-1)(2n-1)$.  The best way to see this is that if we 
define $\xi^{i}_{\pm}= (\theta^{i}\mp\thetatil^{i})$ then the term 
containing $n_{ij}$ in $\gamma$ is basically 
$n_{ij}\xi^{i}_{+}\wedge\xi^{j}_{+}$.  If the components of the 
covariant derivatives of $n_{ij}$ in this basis are $n^{\pm}_{ijk}$ 
then $d(n_{ij}\xi^{i}_{+}\wedge\xi^{i}_{+}) \sim 
n^{+}_{ijk}\xi^{k}_{+}\wedge\xi^{i}_{+}\wedge\xi^{j}_{+} + 
n^{-}_{ijk}\xi^{k}_{-}\wedge\xi^{i}_{+}\wedge\xi^{j}_{+} \ldots$.  The 
stuff in ellipsis does not involves derivatives of $n_{ij}$.  Since 
$n^{+}_{ijk}$ is linearly independent of $n^{-}_{ijk}$ we see that the 
number of equations we get is $\frac{1}{3!}n(n-1)(n-2) + 
n\times\frac{1}{2}n(n-1)$.  The first remark we make is that the PDEs 
given by (\ref{eq:covdernpsi}) generally make an overdetermined 
system if $n > 1$.  The reason is that there are 
$\frac{1}{3}n(n-1)(2n-1)$ equations for $\frac{1}{2}n(n-1)$ functions 
$n_{ij}$.  This means that for a solution to exist integrability 
conditions arising from $d^{2}n_{ij}=0$ must be satisfied.

Let $t_{ijk}=-t_{jik}$ be a tensor in $(\bigwedge^{2}V)\otimes V$ for 
some $n$ dimensional vector space $V$ with inner product.  The vector 
space $(\bigwedge^{2}V)\otimes V$ has an orthogonal decomposition into 
$(\bigwedge^{3}V) \oplus ((\bigwedge^{2}V)\otimes V)_{\rm mixed}$ where 
the latter are the tensors of mixed symmetry under the permutation 
group.  The orthogonal projectors $\antisym$ (antisymmetrization) and 
$\mixed$ (mixed) that respectively project onto $\bigwedge^{3}V$ and 
$((\bigwedge^{2}V)\otimes V)_{\rm mixed}$ are
\begin{eqnarray}
    (\antisym t)_{ijk} & = & \frac{1}{3}(t_{ijk}+t_{jki}+t_{kij})\;,
    \label{eq:defA}  \\
    (\mixed t)_{ijk} & = & \frac{1}{3}(2t_{ijk}-t_{jki}-t_{kij})\;.
    \label{eq:defM}
\end{eqnarray}
A detailed analysis (see below) of equations
(\ref{eq:Hnppsi}), 
(\ref{eq:Htilnpppsi}), (\ref{eq:fnpsi}) and (\ref{eq:ftilnpsi}) shows 
that they determine $\antisym n'$, $\antisym n''$ 
and $\mixed(n'+n'')$. These equations do not provide information 
about $\mixed(n'-n'')$.

To solve the equations above it is best on introduce the following 
auxiliary tensors:  
\begin{eqnarray}
    V_{ijk} &=& H_{ijk}
    -(f_{lij}n_{lk}+f_{ljk}n_{li}+f_{lki}n_{lj}) \;,
    \label{eq:defV} \\
    \Vtil_{ijk} &=& \Htil_{ijk}
    +(\ftil_{lij}n_{lk}+\ftil_{ljk}n_{li}+\ftil_{lki}n_{lj}) \;,
    \label{eq:defVtil} \\
    W_{ijk} &=& (f_{kij}-n_{lk}f_{lij}) = m_{kl}f_{lij}\;,
    \label{eq:defW} \\
    \Wtil_{ijk} &=& 
    (\ftil_{kij}+n_{lk}\ftil_{lij})=\ftil_{lij}m_{lk}\;.
    \label{eq:defWtil}
\end{eqnarray}
They are all skew symmetric under the interchange $i\leftrightarrow j$
and $V,\Vtil$ are totally antisymmetric.  Given a value for $n_{ij}$,
these tensor are determined by the geometric data which specifies the
sigma models.  This data is not independent because these tensors are
linearly related due to the right hand sides of (\ref{eq:Hnppsi}),
(\ref{eq:Htilnpppsi}), (\ref{eq:fnpsi}) and (\ref{eq:ftilnpsi}). 

A little algebra shows that
\begin{equation}
    n'+n''=W-V=-\Wtil+\Vtil\;. 
    \label{eq:npplusnpp}
\end{equation}
All the content of (\ref{eq:Hnppsi}), (\ref{eq:Htilnpppsi}), 
(\ref{eq:fnpsi}) and (\ref{eq:ftilnpsi}) is contained in 
(\ref{eq:Hnppsi}), (\ref{eq:Htilnpppsi}) and (\ref{eq:npplusnpp}). 
These equations place constraints on $V,\Vtil,W,\Wtil$.   
Immediate conclusions are that
\begin{eqnarray}
    W+\Wtil & = & V+\Vtil\;, 
    \label{eq:WplusWtil} \\
    \mixed(W+\Wtil) & = & 0 \;,\\
    \antisym W & = & \frac{2}{3}V + \frac{1}{3}\Vtil\;,
    \label{eq:WV}  \\
    \antisym \Wtil & = & \frac{1}{3}V + \frac{2}{3}\Vtil\;.
    \label{eq:WVtil}
\end{eqnarray}
In deriving the last two equation we used (\ref{eq:Hnppsi}),
(\ref{eq:Htilnpppsi}) and applied the $\antisym$ operator to
(\ref{eq:npplusnpp}).  The equations above imply linear algebraic
relations among the data that defines the sigma models.  They
tell us that there exists a tensor $U_{ijk}$ of mixed
symmetry, \emph{i.e.}, $U_{ijk}=-U_{jik}$ and $\antisym U=0$ such that
\begin{eqnarray}
    W & = & +U + \frac{2}{3}V + \frac{1}{3}\Vtil\;,
    \label{eq:Wsoln}  \\
    \Wtil & = & -U + \frac{1}{3}V + \frac{2}{3}\Vtil\;.
    \label{eq:Wtilsoln}
\end{eqnarray}

Collating all our information we can now write down the 
$\frac{1}{3}n(n-1)(2n-1)$ first order linear PDEs that determine 
$n_{ij}$:
\begin{eqnarray}
    \antisym n' & = & -\frac{1}{3}V\;,
    \label{eq:PDEnp}  \\
    \antisym n'' & = & +\frac{1}{3}\Vtil\;,
    \label{eq:PDEnpp}  \\
    \mixed(n' + n'') & = & +U\;.
    \label{eq:PDEmixed}
\end{eqnarray}
There is no equation for $\mixed(n' - n'')$.
It is worthwhile to note that
\begin{eqnarray}
    n' & = & \half U + \half \mixed(n'-n'') -\frac{1}{3}V \;,
    \label{eq:npsolution}  \\
    n'' & = & \half U - \half \mixed(n'-n'') +\frac{1}{3}\Vtil \;.
    \label{eq:nppsolution}
\end{eqnarray}

You can envision using this formalism in four basic scenarios.
\begin{enumerate}
    \item  Test to see if two sigma models $(M,g,B)$ and 
    $(\Mtil,\gtil,\Btil)$ are dual to each other. This entails the 
    construction of the symplectic manifold $P$.

    \item Given a sigma model $(M,g,B)$ and a symplectic manifold $P$,
    naturally associated with $M$, can you construct the dual sigma
    model $(\Mtil,\gtil,\Btil)$?
    
    \item \label{item:symplectic}
    Given a symplectic manifold $P$ that admits a bifibration, attempt
    to construct dual sigma models.
    
    \item Find all symplectic manifolds $P$ that admit dual sigma
    models.
\end{enumerate}

\subsection{Covariantly constant $n_{ij}$}
\label{sec:covconst}

Here we show that the assumption of covariantly constant $n_{ij}$
leads to a flat connection on $P$.  Assume that in our adapted
coframes the $n_{ij}$ are covariantly constant with respect to the
$\psi$ connection, \emph{i.e.}, $n'_{ijk}=n''_{ijk}=0$.  In this case
it is immediate from (\ref{eq:fnpsi}) and (\ref{eq:ftilnpsi}) that
$f_{ijk}=\ftil_{ijk}=0$.  Subsequently we see from (\ref{eq:Hnppsi})
and (\ref{eq:Htilnpppsi}) that $H=\Htil=0$.  From (\ref{eq:fppTpp}) we
see that $T''_{ijkl}=0$ and thus the curvature vanishes,
$\Psi_{ij}=0$.  
We are mostly interested in local
properties so we might as well assume $P$ is parallelizable.  We can
use parallel transport with respect to this connection to get a global
framing.  In this special framing the connection coefficients vanish
and thus we can make the substitution $\psi_{ij}=0$ in all the
equations in Section~\ref{sec:psi}.
Note that the orthonormal coframes satisfy
$d\theta^{i}=d\thetatil^{i}=0$ and thus $M$ and $\Mtil$ are manifolds
with cover $\bbR^{n}$.  Following up on remarks made in
Section~\ref{sec:nequals0} we see that this is the case of abelian
duality but with constant $n_{ij}$ in the adapted frames corresponding
to constant $B_{ij}$ and $\Btil_{ij}$.

\subsection{Case of $\tilde{f}_{ijk}=0$.}
\label{sec:ftilzero}

What is the most general manifold $M$ whose dual $\Mtil$ has cover
$\bbR^{n}$?    Note that by (\ref{eq:ftiltoTtil}) we have
that $\Ttil_{ijk}=0$ and thus $T''_{ijlm}=-\Ttil'_{ijml}=0$.  This
means that the curvature (\ref{eq:psicurvature}) of the connection
$\psi_{ij}$ vanishes. Again using the remarks 
just made we can choose a parallel framing such 
that $\psi_{ij}=0$. 
Since $\ftil_{ijk}=0$ we have that $d\thetatil^{i}=0$ and
thus locally there exists functions $\xtil^{i}$ such that
$\thetatil^{i}=d\xtil^{i}$. We also have that $d\theta^{i}=-\half 
f_{ijk}\theta^{j}\wedge\theta^{k}$. Previous arguments also tell us 
that $f_{ijk}$ is the pullback of a tensor on $M$. From (\ref{eq:defVtil}) we see 
that $\Vtil = \Htil$ and from
(\ref{eq:defWtil}) we have 
that $\Wtil=0$. Equation~(\ref{eq:Wtilsoln}) tells us that $U=0$ 
and $V=-2\Vtil=-2\Htil$. Inserting into (\ref{eq:Wsoln}) we find that
\begin{equation}
    W_{ijk}= m_{kl}f_{lij}= (\delta_{kl}+n_{kl}) f_{lij}
    = -\Htil_{ijk}\;.
    \label{eq:mfHtil}
\end{equation}
An elementary consequence of this equation is that if $\Htil=0$ then 
$f_{ijk}=0$ and $M$ is also a manifold with cover $\bbR^{n}$. 
Inserting the above into (\ref{eq:defV}) we find that
\begin{equation}
    \Htil_{ijk}= H_{ijk} 
    -(f_{ijk}+f_{jki} +f_{kij})\,.
    \label{eq:HtilHf}
\end{equation}
The left hand side is the pullback of a tensor on $\Mtil$ and the
right hand side is the pullback of a tensor on $M$ thus each side must
be constant.  We have learned that $\Htil_{ijk}$ are constants.  We
assume that $n_{ij}$ are not constant (see
Section~\ref{sec:covconst}).  From (\ref{eq:mfHtil}) we
expect the $f_{ijk}$ not to be constant. Let $\mu_{ijk}$ be a tensor 
of mixed symmetry then from (\ref{eq:npsolution}) and 
(\ref{eq:nppsolution}) we see that
\begin{equation}
    dn_{kl} = \left(\mu_{klm} -\frac{2}{3}\Htil_{klm}\right)\theta^{m}
       + \left(-\mu_{klm} + \frac{1}{3}\Htil_{klm}\right)\thetatil^{m}\;.
    \label{eq:dnder}
\end{equation}
The answer to the question, ``What is the most general manifold $M$ whose
dual $\Mtil$ has cover $\bbR^{n}$?'' and the construction of the 
duality transformation is given by the general solution to the following 
system of exterior differential equations:
\begin{eqnarray}
    (\delta_{kl}+n_{kl}) d\theta^{l} & = & \half 
    \Htil_{kij}\theta^{i}\wedge\theta^{j}\;,
    \label{eq:ftil1}  \\
    d\thetatil^{i} & = & 0\;,
    \label{eq:ftil2}  \\
    dn_{kl} &=& \left(\mu_{klm} -\frac{2}{3}\Htil_{klm}\right)\theta^{m}
       + \left(-\mu_{klm} + \frac{1}{3}\Htil_{klm}\right)\thetatil^{m}\;.
    \label{eq:ftil3}
\end{eqnarray}

As an example consider the special case of $\Htil=0$.  From
(\ref{eq:mfHtil}) we have that $f_{ijk}=0$.  We are now asking,
``What is the most general duality transformation between manifolds
with cover $\bbR^{n}$?''  The equations above tell us that there
exists functions $x^{i}$ and $\xtil^{j}$ such that $\theta^{i}=dx^{i}$
and $\thetatil^{j}=d\xtil^{j}$.  Equation (\ref{eq:ftil3}) becomes
$dn_{kl}=\mu_{klm}dy^{m}$ where $y^{i}=x^{i}-\xtil^{i}$.  We learn
that $n_{ij}$ is a function of $y$ only.  Since the tensor $\mu$ has
mixed symmetry we see that $d(n_{ij}(y)dy^{i}\wedge dy^{j})=0$ and
thus we conclude that locally there exists functions $r_{i}$ of the
independent variables $y^{j}$ such that
$$
    \half n_{ij}(y)dy^{i}\wedge dy^{j} =
    d\left(r_{i}(y) dy^{i}\right)\;.
$$
We now have all the information required to construct the duality 
transformation. The duality transformations are given by
\begin{eqnarray*}
    \pitil  +\xstil{\relax} & = & \pi +\xs{\relax}\;,
      \\
    \pitil  -\xstil{\relax} & = & 
    -T_{-}(x-\xtil)\left[\pi -\xs{\relax}\right]\;
\end{eqnarray*}
with $T_{-}(y)=(I+n(y))(I-n(y))^{-1}$. By taking the sum and 
difference of the equations above one gets ODEs that can be solved 
for $(\xtil(\sigma),\pitil(\sigma))$ given $(x(\sigma),\pi(\sigma))$.

\section*{Acknowledgments}

I would like to thank O.~Babelon, L.~Baulieu, T.~Curtright,
L.A.~Ferreira, D.~Freed, S.~Kaliman, C-H~Liu, R.~Nepomechie,
N.~Reshetikhin, J.~S\'{a}nchez Guill\'{e}n, N.~Wallach and P.~Windey
for discussions on a variety of topics.  I would also like to thank
Jack Lee for his \emph{Mathematica} package Ricci that was used to
perform some of the computations.  I am particularly thankful to
R.~Bryant and I.M.~Singer for patiently answering my many questions
about differential geometry.  This work was supported in part by
National Science Foundation grant PHY--9870101.

\bigskip
\appendix
\par\noindent
{\bf\Large Appendices}
\section{Facts about orthogonal groups}
\label{sec:orthogonal}
We collate some basic properties of orthogonal groups in this section. An 
orthogonal matrix in $\Orth(n,n)$ may be written in terms of $n\times n$ 
blocks as
\[
\pmatrix{A & B \cr C & D\cr}
\]
where $A^t A + C^t C = I$, $B^t B + D^t D = I$ and $B^t A + D^t C =0$. A 
matrix in the Lie algebra $\so(n,n)$ may be written as
\[
\pmatrix{a & b \cr c & d\cr}
\]
where $a = -a^t$, $d = -d^t$ and $c = -b^t$.

The group $\OrthQ(n,n)\subset \GL(2n,\bbR)$ is defined to be the linear 
transformations which leave
\[
    Q = \left(
    \begin{array}{cc}
        0 & I_{n}  \\
        I_{n} & 0
    \end{array}
    \right)\;.
\]
invariant. A matrix in $\OrthQ(n,n)$ may be written in $n\times n$ blocks as
\begin{equation}
	\pmatrix{W&X\cr Y&Z\cr}
	\label{eq:oq-matrix}
\end{equation}
where $W^t Z + Y^t X = I$, $W^t Y + Y^t W = 0$ and $X^t Z + Z^t X =0$. 
It is very important to observe that matrices of the form
\begin{equation}
	\pmatrix{I&0\cr Y&I \cr}
	\label{eq:Y-matrix}
\end{equation}
where $Y = -Y^t$  are in $\OrthQ(n,n)$.
A 
matrix in the Lie algebra $\soQ(n,n)$ may be written as
\begin{equation}
	\pmatrix{w&x\cr y&z\cr}
	\label{eq:soq-lie}
\end{equation}
where $z = -w^t$, $y = -y^t$ and $x = -x^t$.

Conjugating by the orthogonal matrix
\[
   T = 
   {1\over\sqrt{2}}\pmatrix{I_{n}&-I_{n}\cr I_{n}&I_{n}\cr}
\]
leads one to the observation that
\begin{equation}
    T\left(
    \begin{array}{cc}
        0 & I_{n}  \\
        I_{n} & 0
    \end{array}
    \right) T^{-1} = \pmatrix{-I_{n}&0\cr 0& I_{n}\cr}\;.
    \label{eq:conjugation}
\end{equation}
Therefore the group $\OrthQ(n,n)$ is isomorphic to $\Orth(n,n)$.

We will also need to identify the compact group $\Orth(2n) \cap \OrthQ(n,n)$. To 
do this we observe that at the Lie algebra level $\so(2n) \cap
\soQ(n,n)$ is given by matrices of the form
\[
   \pmatrix{a & b \cr b & a \cr}
\]
where $a^t = -a$ and $b^t = -b$. Conjugating by $T$ one sees that
\[
    T \pmatrix{a & b \cr b & a \cr} T^{-1} =
    \pmatrix{a-b & 0 \cr 0 & a+b \cr}\;.
\]
Therefore, the intersection $\so(2n) \cap\soQ(n,n)$ is 
conjugate to $\so(n) \oplus \so(n)$.

\section{On the structure functions}

In a local orthonormal coframe one has the Cartan structural equations
(\ref{eq:cartan1}).  Locally one can always write $d\theta^{i}= -\half
f_{ijk}\theta^{j}\wedge\theta^{k}$ for some ``structure functions''
$f_{ijk}$ which are skew symmetric in $j\leftrightarrow k$.  If we
write the riemannian connection as
$\omega_{ij}=\omega_{ijk}\theta^{k}$ then
\begin{equation}
    \omega_{ijk}=\half(f_{kij}-f_{ijk}+f_{jik})\;.
    \label{eq:rotationcoeff}
\end{equation}
This allows us to reconstruct all the local Riemannian geometry of
the manifold in terms of the structure functions.

\providecommand{\href}[2]{#2}\begingroup\raggedright\endgroup

\end{document}